\documentclass[journal]{IEEEtran}
\usepackage[utf8]{inputenc}
\usepackage{textcomp}
\usepackage{multirow}
\usepackage{graphicx}
\usepackage{amssymb}
\usepackage{amsmath}

\usepackage{cite}

\ifCLASSINFOpdf
\else
\fi
\begin{document}

\title{Single-participant structural connectivity matrices lead to greater accuracy in classification of participants than function in autism in MRI}

\author{Matthew Leming,
        Simon Baron-Cohen, and
        John Suckling

\thanks{

Support was received from the National Institute for Health Research Cambridge Biomedical Research Centre. Matthew Leming is supported by a Gates Cambridge Scholarship from the University of Cambridge. This research has been conducted using the UK Biobank Resource [project ID 20904], co-funded by the NIHR Cambridge Biomedical Research Centre and a Marmaduke Sheild grant to Richard A.I. Bethlehem and Varun Warrier. The views expressed are those of the author(s) and not necessarily those of the NHS, the NIHR or the Department of Health and Social Care. We would also like to thank NDAR, ABCD, ABIDE I, ABIDE II, and Open fMRI for the use of their data; NDAR and ABCD both have their own acknowledgements, which must be excluded for space requirements.}
\thanks{M. Leming, S. Baron-Cohen and J. Suckling are with the Department
of Psychiatry, University of Cambridge, Cambridge,
Cambridgeshire, CB2 0SZ UK (e-mails: ml784@cam.ac.uk, sb205@cam.ac.uk, js369@cam.ac.uk)}
\thanks{}}

\markboth{}%
{Leming \MakeLowercase{\textit{et al.}}: Stochastic encoding of graphs in deep learning allows for complex analysis of gender classificiation in resting-state and task functional brain networks from the UK Biobank}
\maketitle

\begin{abstract}
In this work, we introduce a technique of deriving symmetric connectivity matrices from regional histograms of grey-matter volume estimated from T1-weighted MRIs. We then validated the technique by inputting the connectivity matrices into a convolutional neural network (CNN) to classify between participants with autism and age-, motion-, and intracranial-volume-matched controls from six different databases (29,288 total connectomes, mean age = 30.72, range 0.42-78.00, including 1555 subjects with autism). We compared this method to similar classifications of the same participants using fMRI connectivity matrices as well as univariate estimates of grey-matter volumes. We further applied graph-theoretical metrics on output class activation maps to identify areas of the matrices that the CNN preferentially used to make the classification, focusing particularly on hubs. Our results gave AUROCs of 0.7298 (69.71\% accuracy) when classifying by only structural connectivity, 0.6964 (67.72\% accuracy) when classifying by only functional connectivity, and 0.7037 (66.43\% accuracy) when classifying by univariate grey matter volumes. Combining structural and functional connectivities gave an AUROC of 0.7354 (69.40\% accuracy). Graph analysis of class activation maps revealed no distinguishable network patterns for functional inputs, but did reveal localized differences between groups in bilateral Heschl's gyrus and upper vermis for structural connectivity. This work provides a simple means of feature extraction for inputting large numbers of structural MRIs into machine learning models.\end{abstract}


%
\IEEEpeerreviewmaketitle

\section{Introduction}

\IEEEPARstart{V}{oxel}-based morphometry (VBM) \cite{Whitwell2009} is a means of detecting structural differences in brain anatomy from T1-weighted MRI across groups. In VBM, images are registered to the same coordinate space and segmented into grey matter, white matter, and CSF volumes, before comparisons are made across voxels or groups of voxels using standard statistical tests. Due to its robustness and effectiveness, VBM has enjoyed significant popularity since it was first introduced \cite{Wright1995,Ashburner2000}. Structural covariance networks \cite{Mechelli2005} correlate tissue volumes estimated by VBM in regions across groups of participants to describe relationships that are interpreted as measures of structural integrity or developmental coherence of the brain.

While there have been several cross-sectional findings of structural brain differences in autism \cite{Redcay2005,Stanfield2008,Nickl2012}, these have not been substantiated by a larger-scale analysis \cite{Haar2016}. Indeed, characterizations of brain structure in autism have been inconsistent across studies of small sample sizes, although differences at different ages may explain some of this variation \cite{Chen2011}; for instance, increased amygdala volumes have been reported in children with autism \cite{Sparks2002,Schumann2004}, but not adults \cite{Stanfield2008}. A meta-analysis of VBM studies in autism found disturbance of brain structure in the lateral occipital lobe, the pericentral region, the right medial temporal lobe, the basal ganglia, and proximate to the right parietal operculum \cite{NicklJockschat2012}. Small-scale studies in children with autism have found altered structural covariance in areas associated with sensory, language, and social development. Altered structural covariance has been found between sensory networks, the cerebellum, and the amygdala in autism \cite{Cardon2017}. In children, McAlonan et al \cite{McAlonan2005} found that structural covariance indicated localized reductions within fronto-striatal and parietal networks and decreases in ventral and superior temporal grey matter, suggesting abnormalities in the anatomy and connectivity of limbic–striatal (i.e., social) brain system. Language ability correlated with cortical structure and covariance \cite{Sharda2017}, and associations with language development are further supported by studies showing abnormal development of the Heschl's gyrus \cite{Prigge2013}, an area where functional activation has been associated with development of `inner speech" \cite{Hurlburt2016}. In adults with autism, structural covariance has shown decreased centrality in cortical volume networks \cite{Balardin2015}.

Autism has been consistently associated with differences in brain function \cite{Simas2015,Muller2008}. Efforts to find differences in functional connectivity relative to neurotypical control groups have characterised autism as exhibiting under-connectivity, and thus greater segregation of functional areas \cite{Just2004,Cherkassky2006,Kennedy2008,Assaf2010,Jones2010,Weng2010}. Other studies, mostly of children and adolescents, found evidence of over-connectivity in specific areas of the brains of those with autism \cite{Cerliani2015,Chien2015,Delmonte2013,Martino2011,Nebel2014a,Nebel2014b}, locating hyperconnectivity to the posterior right temporo-parietal junction \cite{Chien2015} and in striatal areas and the pons \cite{Delmonte2013,Martino2011}. One recent review \cite{Hull2017} posited that autism is likely characterised by a mix of hyper- and hypo-connectivity traits.

\subsection{Machine Learning}

Machine learning has found multiple applications to the analysis of brain images in recent years, including pre-processing, segmentation, and diagnostics. Of great interest has been whole-brain phenotypic classification, in which MRI data of two or more phenotypes (such as sexes, or a diseased group and healthy controls) are trained and classified with a machine learning algorithm. Such studies most often include four steps: (1) selection of MRI modality and derived features that are sensitive to the problem at hand ; (2) feature extraction, to reduce data dimensionality; (3) inputting features to train a machine learning model with the selected architecture; and (4) classification and interpretation.

MRI feature extraction is most often performed using techniques previously developed in image analysis, and the specific method is dependent on the selected modality and features. For instance, based on a large body of research and predictable dimensionality reduction \cite{Behrens2007,Friston2011}, it is common to use for classification of functional connectivity matrices \cite{Meszlenyi2017,Kazeminejad2019,Zubaidi2019} representing correlations in time-series between pre-defined regions derived from blood oxygenation level-dependent (BOLD) sensitive fMRI. Likewise, to classify diffusion weighted images (DWI) it is common to use structural connectivity matrices representing the number of white matter tracts traversing the brain between specific regions \cite{Dodonova2016,Kawahara2017,FrauPascual2019}.

However, while there exists several consensus methods for deriving connectivities from fMRI \cite{Friston2011,Patel2016} and DWI \cite{Behrens2007} (though this is still an active area of research \cite{Seidlitz2018,Paquola2019}), there is no published analogous means of connectivity-based dimensionality reduction for T1-weighted structural MRI, even though it is the most common \cite{mriwebpage} modality available to study. One reason for the lack of common methodology is that reductions from three-dimensional data to network representations with meaningful physiological interpretation are more difficult to produce than reductions of four-dimensional data. In most existing feature extraction methods for T1-weighted MRI, extracted features are typically independent, univariate measures from regions of interest, such as cortical thickness and surface curvature. However, the lack of a connectivity metric leads not only to the loss of spatial encoding seen in network representations, but fewer features overall (i.e., for $N$ ROIs, connectivities output $O(N^2)$ features while univariate measurements output $O(N)$), reducing effectiveness for machine learning.

For this paper, we designed a similarity metric that reduces T1-weighted MRIs to a network representation without an a priori physiological interpretation, then applied it a dataset of autistic individuals and neurotypical controls. We classified these structural connectivity matrices using an ensemble convolutional neural network (CNN) \cite{Leming2020b}, then compared the method to classifications that used fMRI connectivity matrices of the same dataset, as well as estimates of grey matter volumes. We applied this method to an extremely large dataset of participants with autism, representing a disorder for which machine learning classification on both functional and structural data has proven difficult \cite{Plitt2015,Katuwal2015,Hensfield2018,Khosla2018}.

\subsection{Machine Learning in Autism}

Previous machine learning studies of autism have achieved classification accuracies that widely varied depending on the modality used, sample size, data quality, selected methods, and diagnostic criteria. A recent study \cite{Hazlett2017} of 106 high-risk infants between 6-12 months linked brain volume overgrowth to the emergence and severity of autism symptoms, using a deep learning algorithm capable of predicting autism with 81\% specificity and 88\% sensitivity using brain surface information. Another study by the same group \cite{Emerson2017} found that autism could be predicted in 59 6-month-old infants with 81.8\% sensitivity using functional imaging. In the general population, efforts in single-participant classification of autism from MRI data have had mixed results \cite{Anderson2011,Barttfeld2012,Nielson2013,Jung2014,Iidaka2015,Plitt2015,Tejwani2017}, with studies rarely exceeding 80\% classification accuracy \cite{Hull2017}. Again, however, this varies substantially by modality and which site data were collected \cite{Katuwal2015}. In a recent study, Eill et al \cite{Eill2019} performed a classification on individuals with autism and neurortypical controls using structural MRI, DWI, and fMRI data, finding that features derived from fMRI provided the highest accuracies with an SVM classifier. They did, however, encounter the issue of fMRI feature extraction simply producing more variables than its structural counterparts, offering the machine learning model more information to work with, although attempts were made to mitigate this issue.

\subsection{Explainable AI}

Besides classification accuracy, an important component to machine learning in scientific discovery is interpretability. The need for explainable machine learning models in a clinical setting has previously been discussed \cite{Gottesman2019}. Clinicians need to fully understand the decision-making process of an automated diagnosis if they are to eventually rely on it. AI models that make a linear, understandable decision-making process are called ``expert systems". These often rely on human-readable information, such as the diagnostic history of an electronic health record. However, such systems would not be capable of making use of more complex datasets that are not always human-readable, such as medical images or genetic records.

Deep learning models have been shown to be capable, at least to a degree, of making sense of complex datasets, in a way that an explainable expert system \cite{Gottesman2019} would not, in applications like whole-brain MRI diagnostics \cite{Kawahara2017,Khosla2018,Leming2020a,Leming2020b}). Unlike expert systems, which rely on linear and human-designed decision-making, deep learning models' decision-making processes are abstracted by their own complexity, a phenomenon generally referred to as the ``black box problem". Because of the need for clinicians to explain their decisions, this would make deep learning models of limited value. There has been great effort in visualising deep learning models in other contexts in the hope of making them explainable. These methods include occlusion, gradient class activation mapping, and activation maximization \cite{Zeiler2013}. While these methods fail to reveal the exact decision-making process used to make classifications, they are capable of showing which parts of the input data are taken into account for the classification. Use of such techniques can make deep learning models more explainable, and thus more useful in an eventual clinical context. But while such methods help explain machine learning models, the full extent of these techniques, and the exact interpretation of any visualization techniques in a scientific context, is still the subject of ongoing research.

\subsection{Experiments}

In the present work, we present a simple method of deriving structural connectivity matrices from T1-weighted MRI. Our method compared the distributions of grey matter in pairs of parcellated areas of T1-weighted MRI. While this method has no specific physiological interpretation, it acted as an effective means of dimensionality reduction that allowed for T1-weighted MRIs to be encoded into a machine learning model. We describe our dataset, including acquisition and preprocessing methods. We validated these data using a machine learning model previously described \cite{Leming2020b} and made comparisons to the classification accuracy using the corresponding fMRI connectivity matrices of the same participants, as well as lower-dimensional data consisting of grey matter volumes averaged within regions. Finally, we used the output class activation maps (CAMs), combined with graph theoretical techniques, to understand which parts of the brain the model focused on, and whether simple linear regression models could spot the same qualities in these data. We further describe a means of combining our structural connectivity matrices with functional connectivity matrices in the same machine learning model to yield improved accuracy.

\section{Materials and Methods}

\subsection{Dataset}

\begin{table*}[t!]
\centering
\begin{tabular}{llllllllllll}
&&&&&Age&&&&Gender&& \\
Collection & \# Subj. & \# Conn. & Rest & Task & Min & Max & Mean & Stdev & Female & Male & Autism \\ \hline
ABCD & 1049 & 5142 & 2296 & 2846 & 0.42 & 11.08 & 10.12 & 0.69 & 2474 & 2668 & 61 \\
ABIDE & 412 & 412 & 412 & 0 & 6.00 & 45.00 & 17.00 & 7.16 & 45 & 367 & 181 \\
ABIDE II & 682 & 717 & 717 & 0 & 5.22 & 55.00 & 14.39 & 7.39 & 169 & 548 & 350 \\
BioBank & 9791 & 9791 & 9791 & 0 & 40.00 & 70.00 & 55.00 & 7.51 & 5178 & 4613 & 4 \\
NDAR & 1050 & 7958 & 5531 & 2427 & 0.58 & 55.83 & 18.71 & 7.80 & 3816 & 4142 & 930 \\
Open fMRI & 1194 & 5268 & 820 & 4448 & 5.89 & 78.00 & 27.12 & 10.24 & 2346 & 2479 & 29 \\
Total & 14178 & 29288 & 19567 & 9721 & 0.42 & 78.00 & 30.72 & -- & 14028 & 14817 & 1555 \\

\end{tabular}
\caption{Statistics for each dataset used.\label{tab:table2_sf}}
\end{table*}

We used a dataset comprised of 29,288 total instances each with a structural MRI and an functional MRI in both task-activated and task-absent (rest) conditions. Note that that instances were acquired from the same participant. In total, 1555 data points were from participants with autism. These data were drawn from six different databases: OPEN fMRI, the UK BioBank, ABIDE I, ABIDE II, NDAR (minus ABCD), and ABCD (Table \ref{tab:table2_sf}). Covariates of age, gender, task were also compiled.

\subsection{Pre-processing and feature extraction}

The full pre-processing pipeline for functional data is described elsewhere \cite{Leming2020a}. Functional data were preprocessed using SpeedyPP. Data were first skull stripped. Motion was regressed from time series using wavelet despiking \cite{Patel2014}. Data were then registered to the stereotatic space of the Montreal Neurological Institute (MNI), after which they were overlaid on the 116-area AAL parcellation. Datasets with greater than 10\% regional dropout, or which otherwise failed the SpeedyPP stage, were excluded. The remaining datasets are presented in Table \ref{tab:table1_sf}. $116\times 116$ functional connectivity matrices were estimated using Pearson correlation on the averaged timeseries within a region. 

To estimate grey matter volumes of each area in the AAL parcellation, structural MRI were first skull stripped using tools from the Analysis of Functional Neuroimages (AFNI) toolbox, then registered to MNI space and grey matter values estimated using FSL VBM. Grey matter volume estimations at each voxel were then averaged within the areas of the AAL parcellation, producing a $116\times 1$ array.

\subsection{Single-Participant Structural Connectivity Matrices}
\begin{figure}
\centering
\includegraphics[width=\columnwidth]{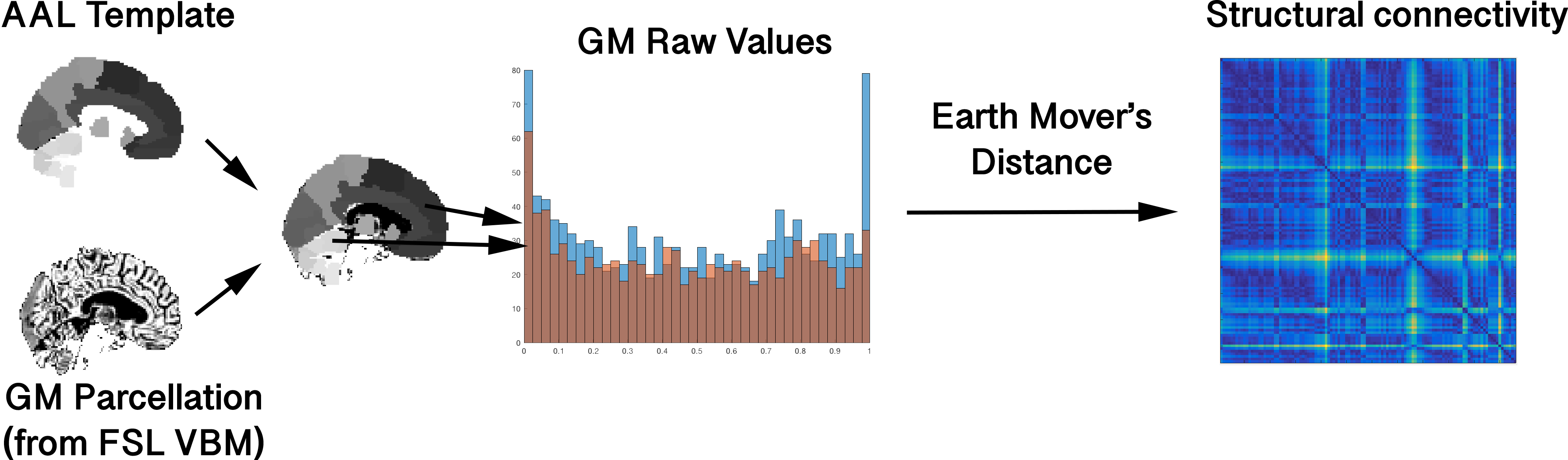}
\caption{Illustration of the procedure used to estimate the structural connectivity matrices used in the present study.}
\label{fig:structural_connectivity_illustration}
\end{figure}

Using the same AAL overlays to derive grey matter volumes, we measured the similarity, $s$ between two regions by comparing the distributions of nonzero voxel values within each region ($u$ and $v$), using the following equation:

\begin{equation}
    s = \underset{\pi \in \Gamma(u,v)}{\inf} \int_{\mathbb{R}\times\mathbb{R}}{|x - y|d\pi(x,y)}
\end{equation}

which is simply the Wasserstein, or Earth-Mover's distance. This is an ideal metric as it non-parametically compares two statistical distributions, regardless of relative region sizes. While this similarity metric does do away with spatial encoding and thus eliminates crucial information such as curvature, it acts as a comparison of the distributions of grey matter volumes between two areas in an easily understood way, and at a low computational cost. An illustration of this is shown in Figure \ref{fig:structural_connectivity_illustration}. While this is a similarity metric that implies no unique physiological relationship between areas, we refer to it as a form of ``connectivity" in line with the commonly used vocabulary in connectomics.

\subsection{Comparison of functional and structural connectivities}

To determine whether functional connectivity and our novel structural connectivity metric shared information in common, we correlated functional matrices from each instance with their corresponding structural matrices, in 10,000 random samples. We then compared these correlations with a null model estimated by correlating random pairings of functional and structural matrices across the collection. This comparison was done by comparing the two sets of 10,000 R values with a t-test, and indicates the amount of common information encoded by both functional and structural connectivities.

\subsection{Machine learning model and training}

We classified individuals with autism and neurotypical controls using, separately, structural connectivity, grey-matter volume, and functional connectivity measurements, as well as a model that combined structural and functional connectivities. We employed the model and training scheme described in \cite{Leming2020b}. This used an ensemble of 300 convolutional neural networks that each scrambled the unique values of input connectivity matrices, losing some spatial encoding information while avoiding biases in output class activation maps (described below).

In building training, test, and validation sets for our models, a multivariate class balancing scheme was used. Equal ratios of autism and neurotypical control participants were enforced, and equal distributions of age, collection, mean framewise displacement of fMRI data, and intracranial volume were maintained in each class. The class balancing scheme divided data into test, training, and validation sets for each model in the ensemble, ensuring participants with multiple functional connectomes were in the same group. Each model was trained on an Adam optimizer for 100 epochs, after which the epoch with the highest accuracy on the validation set was used. This model then made a prediction on each instance in its test set. 

An ensemble of 300 independent CNN models was used to make predictions on the same test set, and an AUROC derived by averaging across instances. When adding models to the ensemble, the AUROC from the aggregated models increased in a predictable way. The AUROCs from between 20 and 300 models were fit to a logarithmic curve with a hard limit in order to predict the projected highest AUROC possible in the limit of a large number of models.

As a result of forced class balancing, each model in the ensemble used an independent subset of approximately 1600 instances. As an effect of this balancing scheme, data from Open fMRI and the UK BioBank, having few participants overall with a diagnosis of autism, were included only infrequently, while data from ABIDE I and II, ABCD, and NDAR were frequently represented.

In total, four cross-sectional classification tasks were undertaken (Table \ref{tab:table1_sf}), specifically: with structural connectivity, with grey matter volumes; with functional encoding; and by combining structural and functional connectivities.

\subsection{Class Activation Map Analysis}

Using the Guided Gradient Class Activation Map (Grad-CAM) algorithm \cite{Selvaraju2017}, which displays areas of the input data most salient in classification, we measured the class activation of each data point in each model proposed, and then averaged these maps generating a $116\times 116$ CAM for both structural and functional connectivity, as well as a $116\times 1$ map for grey matter volume. We correlated the structural and functional CAMs to the measured effect size of differences between autism and neurotypical controls for our connectivity data, as a way to determine the similarity of CAMs to conventional statistics.

Next, we isolated hubs in the $116\times 116$ CAMs. To do so, we first measured the edge betweenness centrality of each edge in our CAMs. We then grouped these values into different communities by maximising modularity of the edge betweenness values (Brain Networks Toolbox \cite{Rubinov2010}). This procedure identified which hubs were most focused on by the classifier.

\section{Results}

\begin{figure}[t!]
\centering
\includegraphics[width=0.45\textwidth]{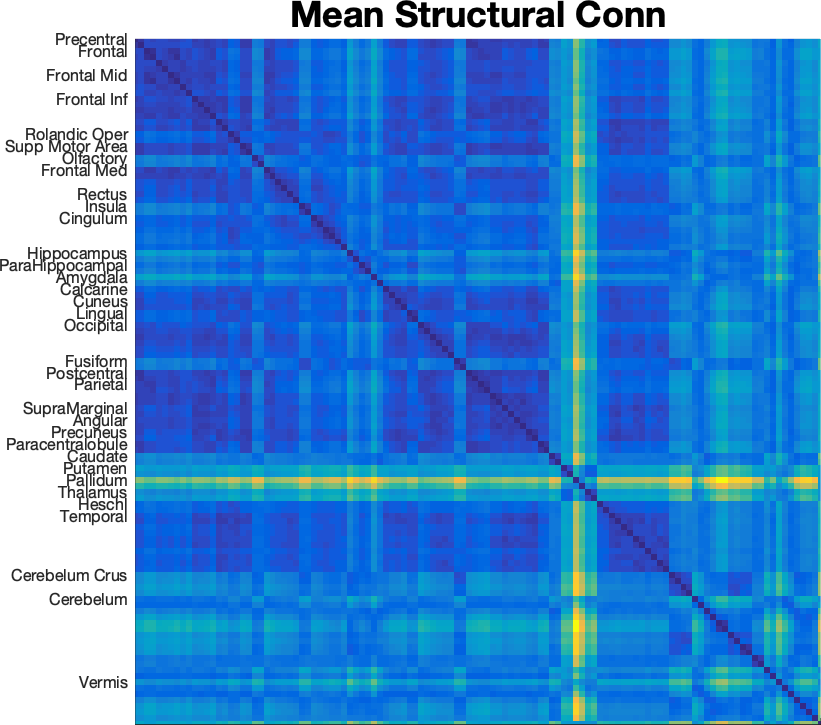}
\\
\includegraphics[width=0.45\textwidth]{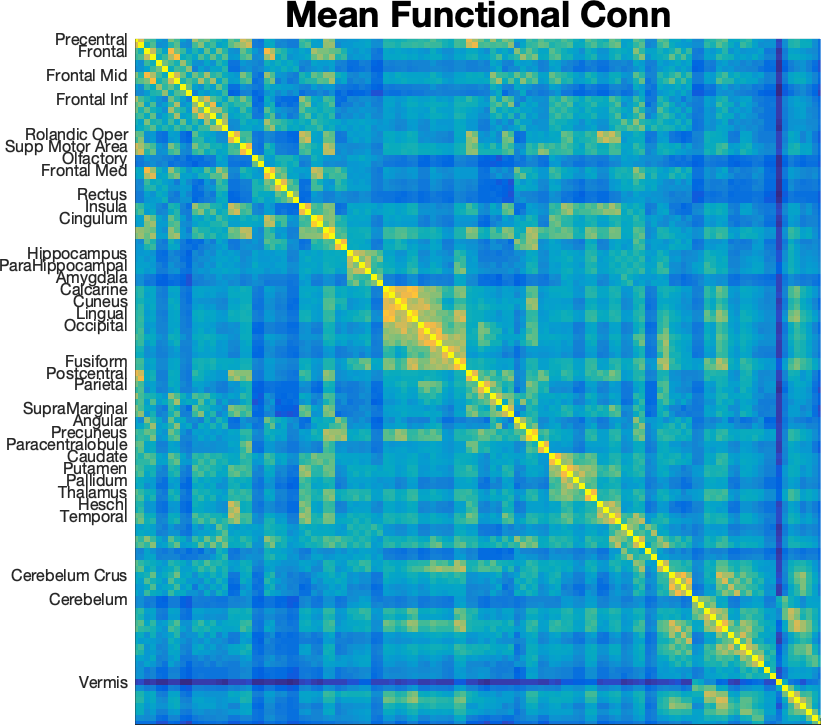}
\\
\includegraphics[width=0.45\textwidth]{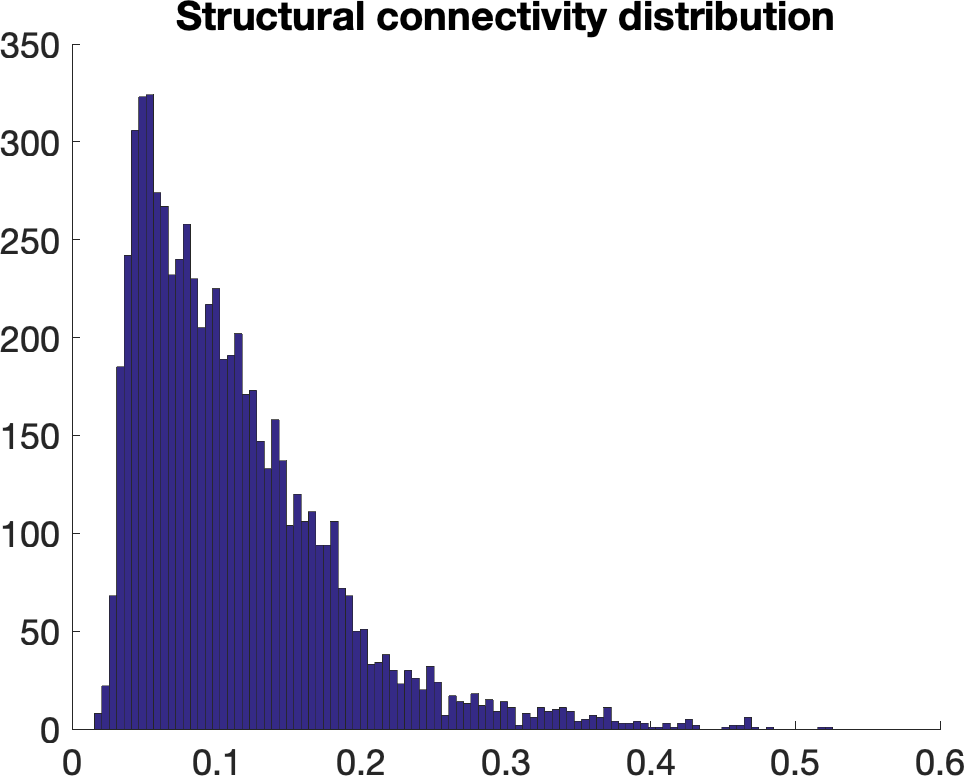}
\caption{The average structural (left) and functional (right) connectivity matrices.}
\label{fig:func_struct_average_comparison}
\end{figure}

\subsection{Comparison of functional and structural connectivities}

Figure \ref{fig:func_struct_average_comparison} shows the average functional and structural connectivity matrices for a balanced group of autism and neurotypical controls. Correlations of functional and structural connectivity matrices from the same participants suggest a modest negative correlation. Across 10,000 random comparisons, the average $R$ value of correlated raw edge values was -0.118 against a null model of -0.108. Subsequent t-tests showed that the $R$ values of the direct comparisons and null model test fell under different distributions (p=$2.216\times 10^{-13}$). This indicates that structural and functional connectivities share only a modest amount of similar information for the same participant.

\subsection{Training}

\begin{table}[t!]
\centering
\begin{tabular}{lll}
              Modality           & AUROC  & Accuracy \\ \hline
Structural conn., Function              & 0.7354 & 69.3980   \\
Structural conn.                        & 0.7298 & 69.7062   \\
Function                         & 0.6964 & 67.7180   \\
Structure (GM vols)                  & 0.7037 & 66.4228   
\end{tabular}
\caption{Respective AUROCs and accuracies of ensemble models on different combinations of data.\label{tab:table1_sf}}
\end{table}

Training accuracies and AUROCs are given in Table \ref{tab:table1_sf}.

\begin{figure}[t!]
\centering
\includegraphics[width=0.45\columnwidth]{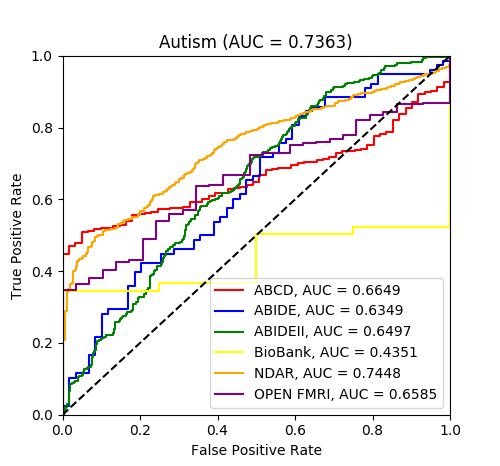}
\includegraphics[width=0.45\columnwidth]{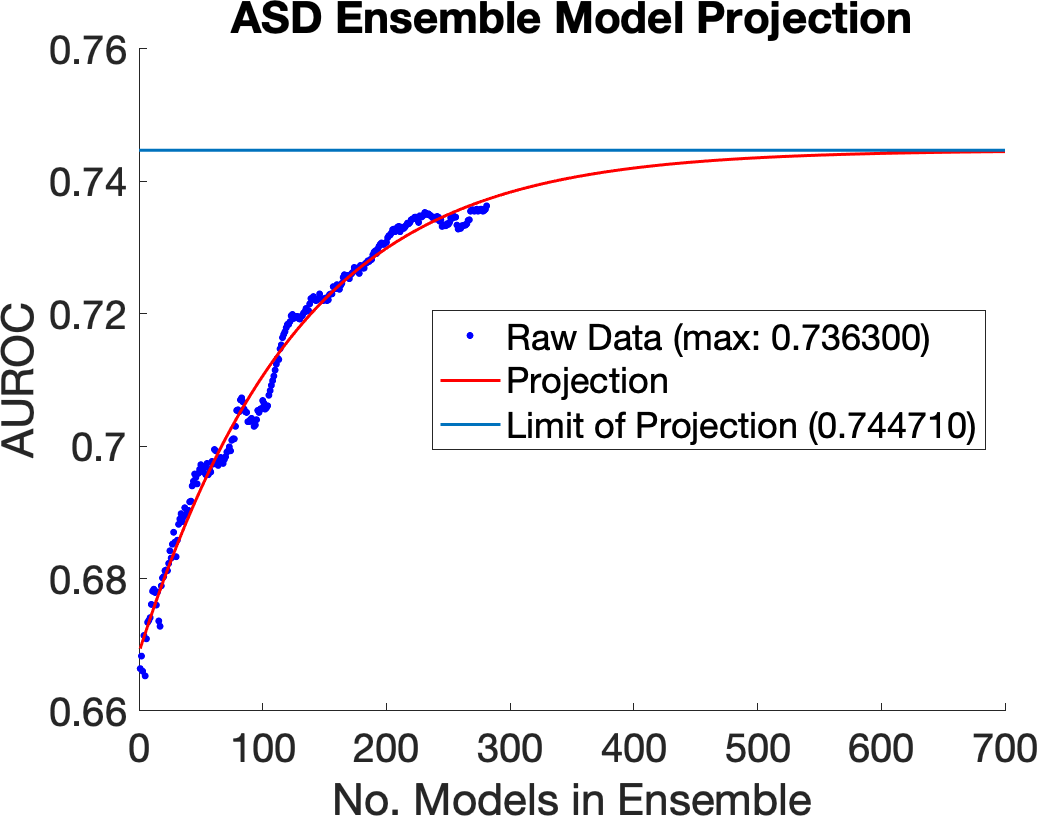}
\caption{Left: Overall AUROCs of each dataset included in the analysis for the structure/function ensemble. Right: Projection of the limit of the structure/function/age ensemble models, given data for ensembles of one to 300 models. Adding independent models \textit{ad infinitum} would result in a maximum predicted AUROC of 0.745.} \label{fig:auroc_plot_asd}
\end{figure}

Classification resulted in a higher AUROC for structural than functional connectivities: 0.7298 and 0.6964, respectively. Classification on univariate grey matter volumes resulted in an AUROC of 0.7037, outperforming functional classification while underperforming structural connectivity classification, although this might be expected considering its lower dimensionality. Combining structure and function resulted in an AUROC of 0.7354 (Figure \ref{fig:auroc_plot_asd}, left), with a projected upper limit of the AUROC of 0.745 (Figure \ref{fig:auroc_plot_asd}, right).

\subsection{Class Activation Map Analysis}

\begin{figure}[t!]
\centering
\includegraphics[width=0.225\textwidth]{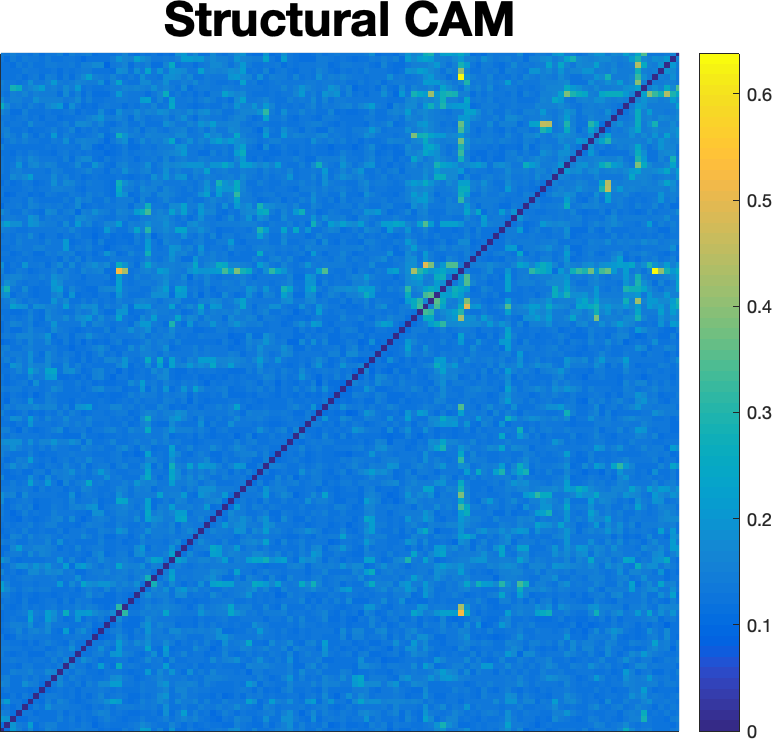}
\includegraphics[width=0.225\textwidth]{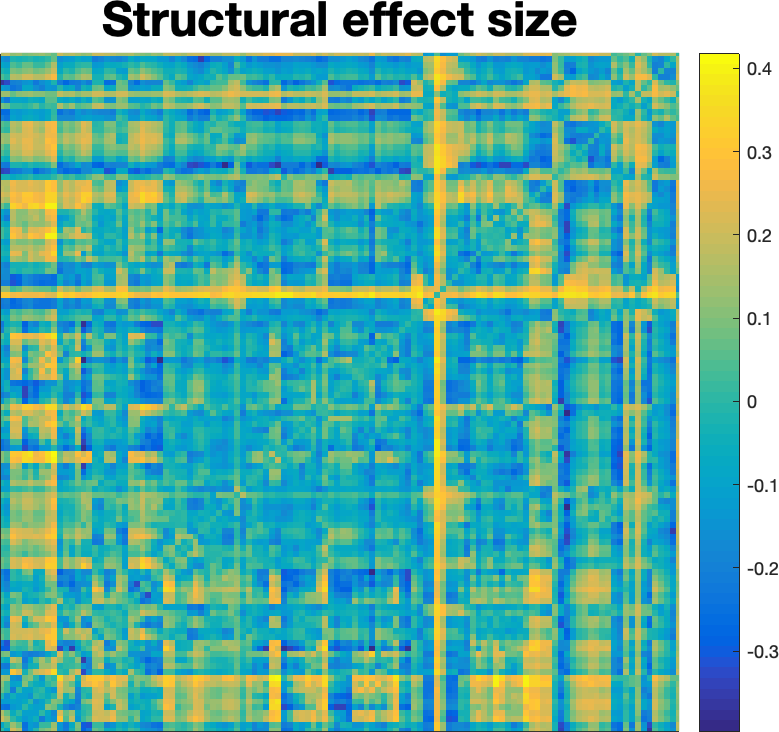}
\includegraphics[width=0.225\textwidth]{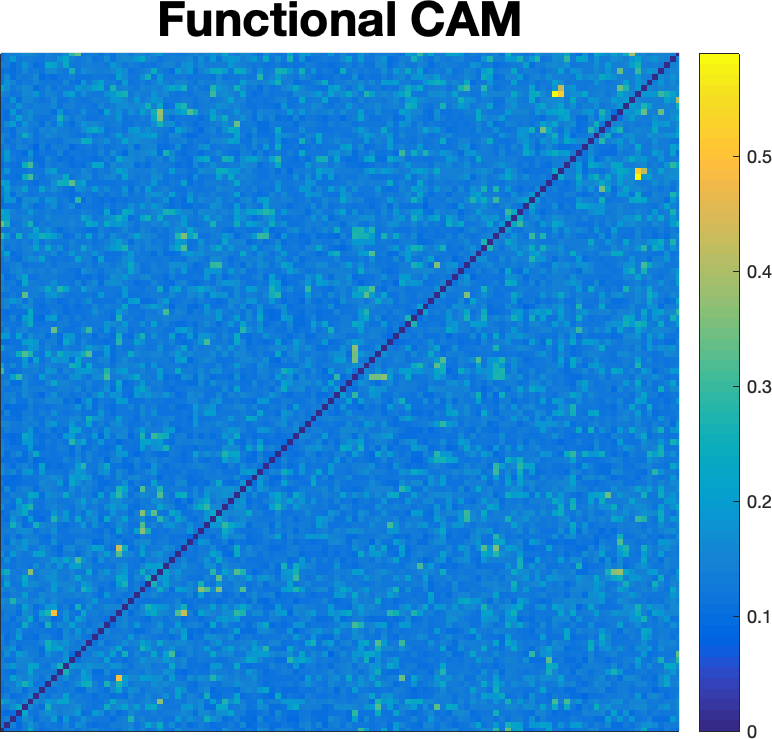}
\includegraphics[width=0.225\textwidth]{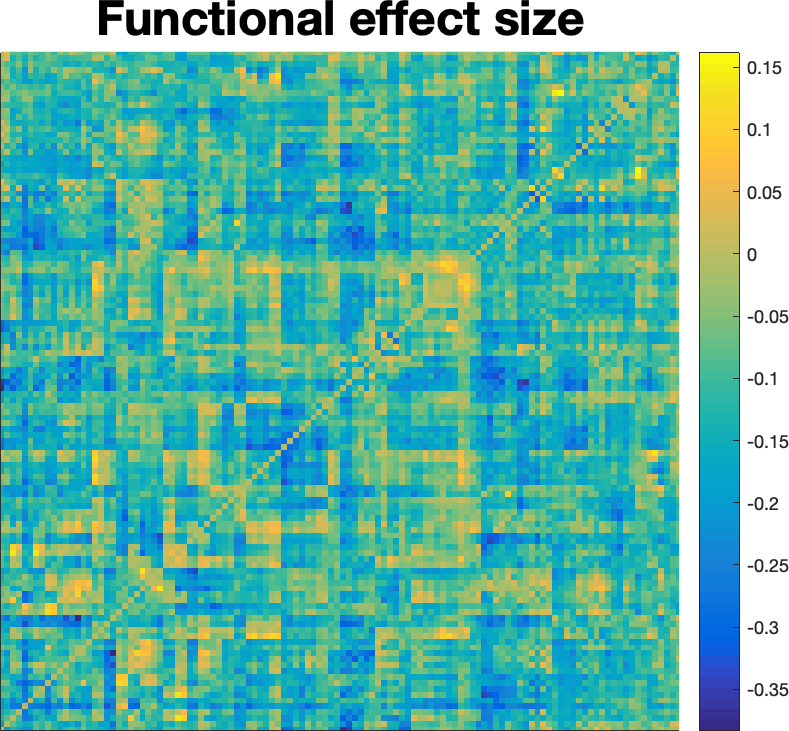}
\caption{A comparison of the effect size of differences between raw matrix values between groups and the averaged class activation maps. Most of the edge differences passed a nonparametric statistical significance test. When comparing the CAM matrix and the effect size matrices using either linear or nonparametric correlation, neither had any statistically significant associations with one another.}
\label{fig:effect_cam_comparison}
\end{figure}

When comparing the output CAMs to their respective functional and structural effect sizes, no statistically significant correlation was observed, and thus the machine learning model relied very little, if at all, on differences detectable by conventional statistics (Figure \ref{fig:effect_cam_comparison}).

\begin{figure*}[t!]
\centering
\includegraphics[width=0.8\textwidth]{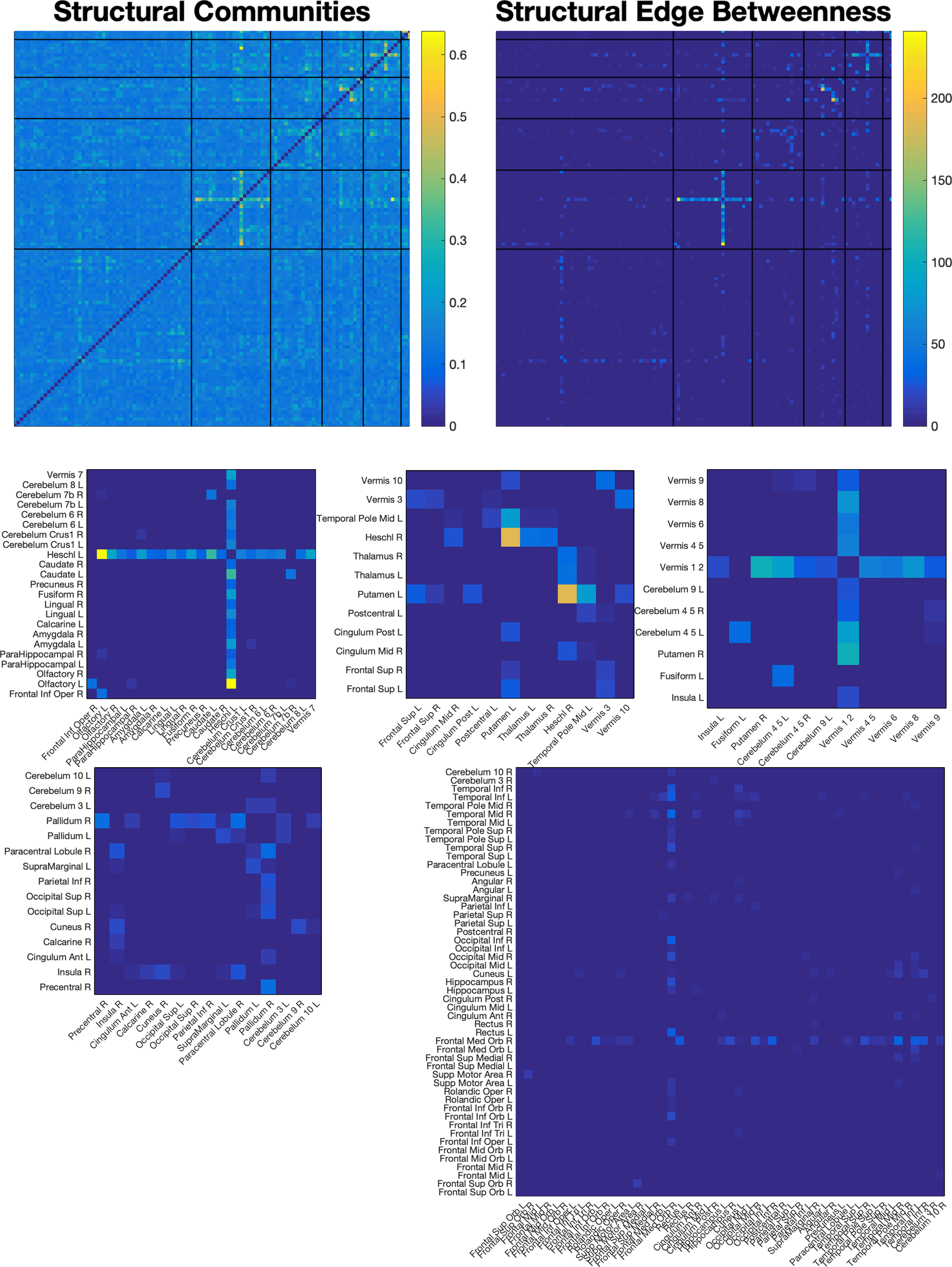}
\caption{The structural hubs targeted by the structure/function/age encoding. Shown here are the class activation maps (upper left) as well as the edge betweenness centralities of the map (upper right), after it as been sorted into six different hubs via modularity maximization. The hubs, with labeled areas, are shown in the bottom half. (Middle) The three most distinct hubs revolve around the left Heschl's gyrus; the right Heschl's gyrus (and, to an extent, the left Putamen); and the upper vermis. The largest hub, in the bottom left, shows scattered-but-weak emphasis on connections to the right frontal medial orbital gyrus. These connections likely reflect the machine learning model's use of comparisons of certain areas to others in order to assess the developmental difference of such areas in autism.}
\label{fig:structural_communities}
\end{figure*}
\begin{figure*}[t!]
\centering
\includegraphics[width=0.8\textwidth]{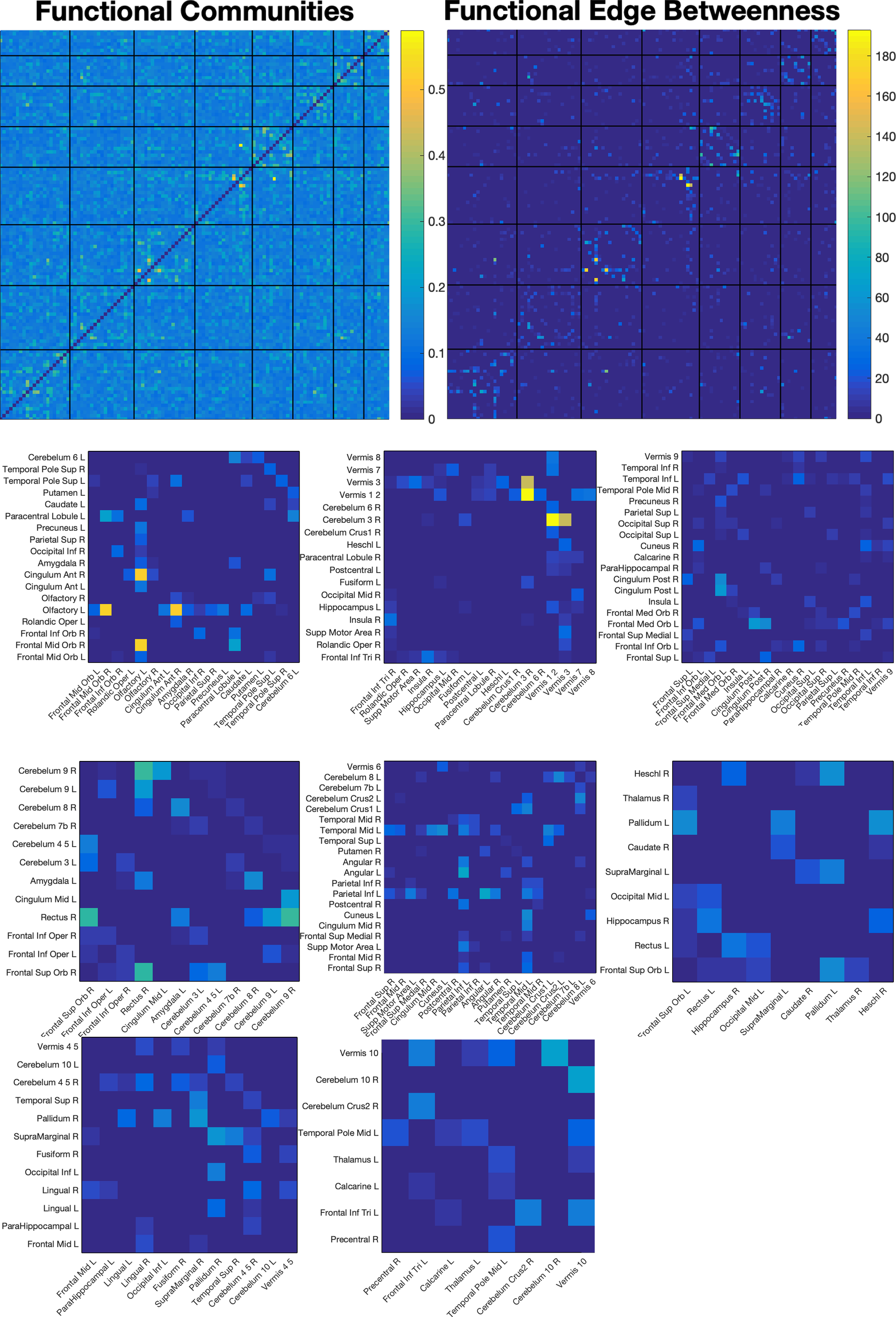}
\caption{Averaged functional class activation maps and the associated edge betweenness centralities, when divided into communities via modularity maximization. Function does not show the same ultra-localized hubness within particular communities in the way that the structural results do, but emphasis is given to several individual connections throughout.}
\label{fig:functional_communities}
\end{figure*}

CAMs for structural and functional connectivities, sorted by different detected communities after edge betweenness centrality was measured, are shown in Figures \ref{fig:structural_communities} and \ref{fig:functional_communities}. Structural CAMs showed five distinct groupings, each with distinct hubs that each centered on one or two localised areas, including the left and right Heschl's gyrus, the upper vermis, the right frontal-medial orbital gyrus, the right pallidum, and the left putamen. The strongest activations were found in left Heshcl's gyrus.

Localisation was also found, though less distinctly, in functional hubs, notably the left inferior parietal lobe, the left middle temporal lobe, the left olfactory bulb, and the upper vermis. However, focus on particular hubs was not a distinctive feature.

\begin{figure}[t!]
\centering
\includegraphics[width=0.4\textwidth]{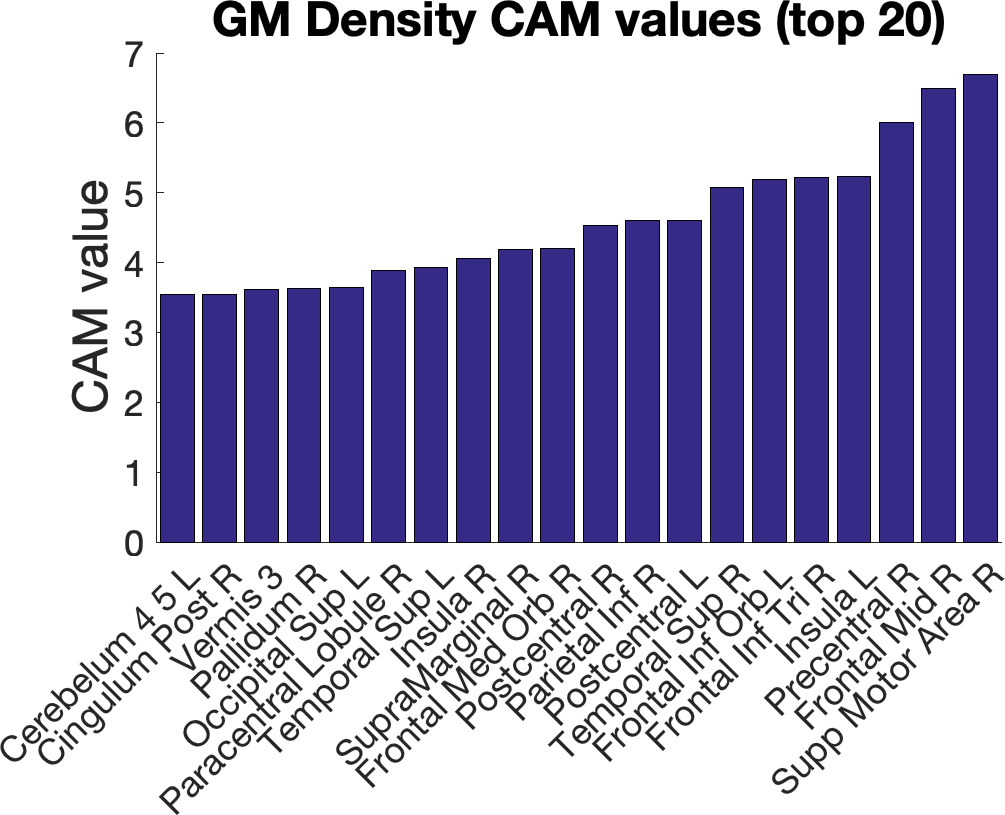}
\caption{Top class activation map value results for the 116-area gray matter density classification, showing the areas most focused on in that classification task. The minimum CAM value (not shown) was 1.3622.}
\label{fig:gm_volume_cam_values}
\end{figure}

CAMs for grey matter volumes are shown in Figure \ref{fig:gm_volume_cam_values}. These results had very little in common with the structural connectivity results, with the strongest five activated areas in the right supplementary motor area, the right middle frontal lobe, the right precentral sulcus, the left insula, and the inferior frontal gyrus triangularis.

\section{Discussion}

This study proposed a new feature extraction method for inputting structural MRIs into a network-based machine learning model, as well as applicable analysis methods to detect areas of that were particularly involved in determining the classification. Estimating single-participant structural connectivity matrices from T1-weighted images without supplementary modalities such as DWI or fMRI is uncommon, and, with few exceptions \cite{Tijms2012} research in this area is relatively undeveloped. In structural covariance, VBM data is used to produce inter-regional relationships at a group level, but this is inapplicable at a single-participant level, which is necessary to make structural MRIs applicable to machine learning models. The proposed method provides a means of doing so and validates it in a practical way.

In the univariate grey matter volume results, the CAMs highlighted the right supplementary motor area, right mid frontal lobe, right precentral sulcus, left insula, right frontal inferior triangularis, left frontal inferior orbital lobe, and the right superior temporal lobe (the top 20 areas are shown in Figure \ref{fig:gm_volume_cam_values}). Comparing the CAM emphasis of the grey matter volumes to the meta-analysis of autism VBM studies in \cite{NicklJockschat2012}, which found six areas with consistently altered grey matter volumes, some similarities can be seen, notably in the right superior temporal lobe where grey- and white-matter volume differences in the right medial temporal lobe and the left post central gyrus.

Functional analysis did not reveal a pattern of local hubness characterising structural  connectivity differences, but rather focused on specific connections. However, a number of general functional communities were identified (Figure \ref{fig:functional_communities}). Meta-analyses of studies in functional connectivity differences associated wtih autism have not found consistent differences in the brain, but rather in network-wide measures \cite{Hull2017}. The lack of hub emphasis in functional results may be additional evidence of network-wide, rather than localized differences between autism and neurotypical control groups seen in other recent findings \cite{Suckling2015}.

In structural connectivity, three definitive hubs were identified: left Heschl's gyrus, right Heschl's gyrus, and the upper vermis. The right pallidum and fronto-medial orbital region also showed relatively strong local hubs, though to a lesser degree. Emphasis of the Heschl's gyrus is in agreement with recent studies in developmental autism, having been implicated previously as an area that develops atypically in autistic children \cite{Prigge2013}. Function of the area has been associated with development of ``inner speech" \cite{Hurlburt2016}, indicating a difference in development of language capabilities. Our findings differ in that they found this emphasis in \textit{structure} and not \textit{function}, but this may be reflective of the lower variability of differences across a single area in the development of grey matter as opposed to function, which likely varies far more across participants, and age groups. The cerebellum, meanwhile, has consistently been cited as an area of difference between individuals with autism and neurotypical controls during development \cite{Chen2011}, as well as an area of difference in structural covariance associated with autism \cite{Cardon2017}.

The structural connectivity CAMs resulting from our study revealed an emphasis on a number of distinct and localised areas, and these areas were clarified by use of an edge centrality measurement combined with modularity maximization to isolate hubs. The edge betweenness step was added by necessity to place extreme emphasis on a smaller number of more central edges, and only then could modularity maximisation isolate hubs in a meaningful way (see Figure \ref{fig:structural_communities}).

Our structural connectivity method's efficacy with the machine learning model suggests that it encoded practically useful information about brain structure, but the interpretation of what these structural hubs indicate physiologically is more complicated. While some correlation is present (Figure \ref{fig:func_struct_average_comparison}) the functional and structural connectivities show largely different patterns. Furthermore, considering that the method used to estimate structural connectivity was a similarity metric, the emphasis on these hubs was less likely an indication that they were centers of a physiological brain network characterising autism. Because the strength of connections was a comparison of grey matter distributions, it is more likely that connections to the identified hubs were used by the machine learning algorithm as a proxy for detecting subtle changes in the morphology of grey matter within those specific regions. Edges connecting to these structural hubs were probably an indirect indication of differences in grey matter between two areas, and the individual connections themselves would not indicate any special physiological relationship. However, this still means that the hubs themselves were important in characterising autism. This lack of an explicit physiological interpretation of our metric, however, does not detract from its utility in the context of machine learning. This structural connectivity metric may simply be viewed as a way of encoding relative spatial information about the morphology of individual areas of the brain.

The univariate grey matter volume results further complicate interpretation because areas different from the structural connectivity results were emphasised by the CAMs, even though both univariate grey matter volumes and structural connectivities were derived from the same imaging data. This brings up three key points. First, the method of encoding data is important because it presents different types of information to the machine learning model. Structural differences in autism (and likely other phenotypic differences) may vary in different ways that are only apparent under specific methods of encoding, and thus the model may have focused on different areas, depending on which method of encoding was performed. This is important for both interpreting the results in the context of a specific machine learning task and understanding the underlying physiological implications. Second, the emphases presented by Grad-CAM were \textit{relative}; that is, in analysing the distribution of Grad-CAM values, we saw that the model took all areas into account (Figure \ref{fig:gm_volume_cam_values}), although with highest focus on the few areas that seemed to hold more influence in the final classification task. This does not, however, mean that other areas were ignored entirely. Third, because of the higher dimensionality of structural connectivities over grey matter values, it may be the case that the machine learning model assumed information about grey matter volumes from a small number of edges, while information about differences in morphology of other areas (e.g., the left and right Heschl's gyrus), which were not present in the univariate feature extractions, required emphasis by a greater number of edges; this may be crucial to understanding differences in autism generally, or it may have simply helped the model increase AUROC by a margin of 0.0891 between the univariate and connectivity classification tasks. Stated informally, differences in morphology detected by our structural connectivity matrices were more subtle, and so they required the emphasis of a larger number of edges.

In developing this method, other means of estimating single-participant connectivity matrices from T1-weighted MRI were considered, such as estimating the correlation between different univariate measurements (cortical thickness, curvatures, and so on) of the structural image \cite{Seidlitz2018,Paquola2019}, but this was too computationally intensive for a large dataset. Another method was investigated that involved finding the difference between group structural covariance matrices with and without a certain participant. While classifications on these matrices were successful, the matrices themselves varied to such an extent that the output CAMs were inconsistent. In the end, the proposed method was used because of its simplicity and effectiveness in classification.

It is notable that none of the classification accuracies presented in this paper approached the success required for a clinical diagnosis, which would need to consistently exceed 95\% accuracy on a substantially large dataset. A likely reason for the comparatively low accuracy in this study specifically is the large dataset size, which, in the context of whole-brain MRI classification, has previously been associated with a drop in accuracy \cite{Katuwal2015,Arbabshirani2017}. Nonetheless, deep learning models are useful in these contexts both as statistical models in and of themselves to study autism, and as building blocks to approach clinical-quality accuracy in the future.

Finally, we combined our structural connectivity metric with functional connectivity raising the final AUROC. This shows that our method does not have to be considered as a replacement for any previous methods, but may be used in combination with them in order to make single-participant classifications more effective.

\section{Conclusion}

The present work offers a means of encoding T1-weighted MRI for use in network-based machine learning models, and with a machine learning classification task we have demonstrated an increase in accuracy in classifying individuals with autism when compared with both functional connectivities and classification of univariate grey matter volumes. Furthermore we presented methods of identified areas emphasized by the machine learning model, demonstrating the importance of data encoding and highlighting complications with interpreting results when the feature extractions have no specific physiological interpretation. While this tradeoff, interpretability for higher accuracy, will likely continue to be an issue in machine learning with scientific data, the effects of data encoding on accuracy point towards feature extraction methods as a future direction of investigation. 


%




\ifCLASSOPTIONcaptionsoff
  \newpage
\fi

\bibliographystyle{IEEEtran}
\bibliography{IEEEabrv,references}
\end{document}